\documentclass[prl,showpacs,showkeys,nofootinbib,twocolumn,floatfix]{revtex4}

\usepackage{graphicx}  %
\usepackage{amsmath}
\usepackage{bm}  %
\usepackage{ulem}

\newcommand{\im}[1]{\mathrm{Im}\left(#1\right)}
\newcommand{\re}[1]{\mathrm{Re}\left(#1\right)}

\newcommand{\omb}{E_D}
\newcommand{\om}{\omega}

\newcommand{\abar}{\bar a}

\newcommand{\Bbar}{\bar{B}}
\newcommand{\Btil}{\widetilde{B}}
\newcommand{\btil}{\tilde{b}}
\newcommand{\mbf}[1]{\bm{#1}}
\newcommand{\dvec}[1]{\frac{\mathrm{d}\mbf{#1}}{(2\pi)^3}\,}

\begin{document}
 
\title{
Inducing Resonant Interactions in Ultracold Atoms \\with a Modulated Magnetic Field}

\author{D.~Hudson Smith}
\email{smith.7991@osu.edu}
\affiliation{Department of Physics,
         The Ohio State University, Columbus, OH\ 43210, USA\\}

\date{\today}

\begin{abstract}
In systems of ultracold atoms, pairwise interactions can be resonantly enhanced by a new mechanism which does not rely upon a magnetic Feshbach resonance. In this mechanism, interactions are controlled by tuning the frequency of an oscillating parallel component of the magnetic field close to the transition frequency between the scattering atoms and a two-atom bound state. The real part of the resulting s-wave scattering length $a$ is resonantly enhanced when the oscillation frequency is close to the transition frequency. The resonance parameters can be controlled by varying the amplitude of the oscillating field. The amplitude also controls the imaginary part of $a$ which arises because the oscillating field converts atom pairs into molecules. The real part of $a$ can be made much larger than the background scattering length without introducing catastrophic atom losses from the imaginary part. For the case of a shallow bound state in the scattering channel, the dimensionless resonance parameters are universal functions of the dimensionless oscillation amplitude.
\end{abstract}

\smallskip
\pacs{31.15.-p, 34.50.-s, 67.85.-d}
\keywords{Ultracold atoms, Feshbach resonance, scattering of atoms, molecule formation}
\maketitle

{\it Introduction.---} A unique feature of ultracold atomic physics is the ability to precisely control the interactions among particles all the way from zero interactions to infinitely attractive or repulsive interactions. This tunability has led to many breakthroughs in few- and many-body physics. In most current experiments, interactions are controlled by exploiting a {\it magnetic Feshbach resonance} (MFR), where an external magnetic field is tuned near the value $B_0$ where a pair of unbound atoms becomes degenerate with a two-atom bound state \cite{Review_FR}. For ultracold atoms, the strength of interactions is determined by the s-wave scattering length $a$. Near $B_0$, $a$ is a simple function of the magnetic field $B$:
\begin{equation}\label{eq:feshbach}
\frac{1}{a(B)}=\frac{1}{a_{\mathrm{bg}}}\,\frac{B-B_0}{B-B_0-\Delta}+i\gamma,
\end{equation}
where $a_{\mathrm{bg}}$ is the background scattering length, $\Delta$ is the width of the resonance, and $\gamma$ is non-zero only if the colliding atoms have a spin-relaxation scattering channel. 

Other mechanisms for resonantly enhancing $a$ have been proposed. In {\it optical Feshbach resonance} (OFR) \cite{Shlyapnikov,Julienne_1999,Lett_2000,Grimm_2004,Grimm_2005}, laser light that is slightly detuned from a transition to an electronically excited p-wave molecule induces a resonance in $a$. The resonance properties depend on the intensity of the laser. This technique has major limitations for alkali-metal atoms because the rapid spontaneous decay of the resonance molecule results in inelastic losses and severely limits the maximum enhancement of $a$. In {\it radio-frequency Feshbach resonance} (rfFR) \cite{Julienne_rfFR_2009,Schmiedmayer_rfFR_2010} and {\it microwave Feshbach resonance} (mwFR) \cite{Dalibard_mwFR_2010}, an oscillating magnetic field that is perpendicular to the spin-quantization axis of the atoms couples an atom pair to a molecule in another hyperfine channel, thereby modifying or inducing a resonance in $a$. These methods allow some control over $a$ without introducing dramatic atom loss. One disadvantage of rf/mwFR is that the coupling between an atom pair and the resonance molecule tends to be small, leading to very small enhancement of the real part of $a$. Also, it is difficult to produce large-amplitude rf and mw fields.

In this letter, we examine a new mechanism, {\it modulated-magnetic Feshbach resonance} (MMFR), for resonantly enhancing $a$ in ultracold gases. This mechanism is related to {\it modulated-magnetic spectroscopy} or {\it wiggle spectroscopy}, which has been used to measure molecular binding energies and other properties for several alkali-metal atoms \cite{Wieman2005,Inguscio0808,Chin_2009,Khaykovich1201,Hulet1302}. A MMFR is induced by an oscillating magnetic field, 
\begin{equation}
B(t)=\Bbar + \Btil\cos(\om t)
\label{eq:BofT}
\end{equation}
parallel to the spin-quantization axis of the atoms and near the transition frequency between an atom pair and a bound state either in the scattering channel or another hyperfine channel. Since the magnetic field is longitudinally polarized, the scattering channel and the bound-state channel must have the same $z$-projection of total hyperfine spin. Near a MFR, the transition frequency can be decreased by adjusting $\Bbar$ to bring the resonance molecule closer to the threshold. The frequencies needed for MMFR are much lower than for rf/mwFR. This allows for larger amplitudes of the oscillating magnetic field. Also, because of the parallel polarization of the oscillating field, the coupling between an atom pair and the bound state can be much stronger than for rf/mwFR. Larger oscillation amplitudes combined with stronger coupling results in greater enhancement of the $a$.

We will show that when the applied frequency $\omega$ is near the transition frequency $\omb$, $a$ is a simple function of $\omega$:
\begin{equation}\label{eq:resonance_eqn}
\frac{1}{a(\om)}=\frac{1}{\abar}\,\frac{\om-\om_0}{\om-\om_0-\delta}+i\gamma,
\end{equation}
where $\abar$ is the scattering length in the absence of the modulated field, $\omega_0$ is close to $\omb$, and $\delta$ is the width of the resonance. The inverse scattering length has been given a frequency-independent positive imaginary part $\gamma$ that is only important very close to the resonance. The parametrization in Eq.~\eqref{eq:resonance_eqn} ensures that $\im{a}\leq 0$ for all $\omega$, as required by unitarity. The maximum value of $|\re{a}|$ is $1/2\gamma$. We will show that MMFR can give large enhancements to $\re{a}$ while still having small $\im{a}$.

The imaginary part of $a$ in MMFR arises from collisions in which a pair of low-energy atoms emits one or more quanta of frequency $\omega$ and forms a molecule. A complex $a$ also arises when controlling the scattering length of $^{85}$Rb with MFR because the only accessible broad resonance occurs in a hyperfine configuration with a spin-relaxation channel. However, this has not prevented pioneering studies of few- and many-body physics using $^{85}$Rb atoms \cite{Wieman_2002,Wieman_2005,Robins_2011,Jin_2012,Cornish_2013,Jin_2014}. In OFR, a complex $a$ arises because of spontaneous and stimulated decay of the excited p-wave molecule, and the maximum enhancement to $a$ is proportional to the laser intensity \cite{Julienne_1999}. In contrast, the bound state in MMFR can be stable with respect to spontaneous decay, and the maximum enhancement to $a$ is inversely proportional $\gamma$ which is proportional to the square of the amplitude of the oscillating field as we will show. Thus, MMFR may provide more viable applications for alkali-metal atoms than OFR. Furthermore, experimental control of the oscillation frequency of magnetic fields tends to be much better than control over the dc value of the magnetic field. Thus MMFR could be used to more precisely tune $a$ in the vicinity of narrow MFRs.

In this letter, we introduce a general formalism for treating the scattering of neutral atoms from short-range, time-periodic potentials. We then use this formalism to calculate $a(\omega)$ analytically in the zero range limit in the case that the resonance molecule is a shallow dimer with energy $E_D=1/(m\abar^2)$. We show the existence of a scattering resonance as a function of $\omega$ near $E_D$, and we analytically extract the dependencies of the resonance parameters $\delta$, $\Delta\om_0=\om_0-\omb$, and $\gamma$ on $\Btil$. For small $\Btil$, the resonance parameters each scale as $\Btil^2$. The dimensionless resonance parameters $\delta/\omb$, $\Delta\om_0/\omb$, and $\gamma\abar$ are universal functions of a dimensionless variable proportional to $\Btil$. These analytic results are confirmed numerically by a toy model with a square-well potential with oscillating depth. We conclude by discussing a possible experimental application of MMFR.

{\it Scattering from a time-periodic potential.---} 
Floquet theory provides a natural framework for treating the time-evolution of wavefunctions in the presence of time-periodic potentials. In this framework, the problem of electrons scattering from neutral atoms in the presence of laser fields has been thoroughly treated, including electron-atom frequency-controlled scattering resonances (for a pedagogical overview, see \cite{Joachain:book:2014}). The primary difference between that problem and the scattering of neutral atoms in an oscillating magnetic field is that in the latter case, the effect of the oscillating field is confined within a short range. Specifically, we consider the s-wave scattering of neutral atoms with a modulated instantaneousness scattering length $a(t)$ being controlled, e.g., with the time-dependent magnetic field $B(t)$ in Eq.~\eqref{eq:BofT}. Since we are primarily interested in resonance phenomena, neither the time-independent nor the time-periodic part of the potential can be treated perturbatively. This disallows a simple extension of the perturbative approach introduced by Langmack, Smith, and Braaten \cite{Langmack:prl:2015}.

Floquet's theorem asserts that an incident particle flux with energy $k^2/m$, where $\mbf{k}$ is the relative momentum of the scatterers, couples to the so-called {\it Floquet modes} with energies $k^2/m+n\omega$, where $n$ can be any integer. This motivates a coupled-channels treatment of the scattering problem where the different channels correspond to the incoming channel as well as the Floquet modes. The scattering state $|\psi(t)\rangle$ is the solution to the time-dependent Schr\"odinger equation with a time-periodic potential $V(t)=\sum_{n} V_n\exp{(-in\omega t)}$ along with the boundary condition that $|\psi(t)\rangle$ approaches the free incoming state $|\phi(t)\rangle$ as $t\rightarrow -\infty$. We assume the asymptotic states can be approximated as a momentum eigenstate as for scattering from time-independent potentials. The validity of this assumption depends on the strength and frequency of the oscillating field: if the field is too strong or too slow, the coupled-channels approach will fail. Fortunately our approach has a built in ``safety mechanism'' that will signal the breakdown of this approximation.

The coupled-channels equations which determine the Floquet components of the wavefunction are extracted by inserting for $|\psi(t)\rangle$ and $|\phi(t)\rangle$ their Fourier expansions in terms of their frequency representations $|\psi(\omega)\rangle$ and $|\phi(\omega)\rangle$. This leads to an infinite, coupled set of Lippmann-Schwinger equations:
\begin{equation}
|\psi_n\rangle = |\mbf{k}\rangle\delta_{n,0} + \frac{1}{k_n^2/m-H_0 + i\epsilon}\sum\limits_{s=-\infty}^\infty V_{n-s}|\psi_s\rangle,
\label{eq:LSEPsi}
\end{equation}
where $H_0$ is the kinetic energy operator and we have defined $k_n =\sqrt{k^2+n\omega}$ and $|\psi_n\rangle=|\psi(k_n^2/m)\rangle$.
Projecting onto coordinate space and examining the wavefunction in the region far outside the scattering potential, we find
\begin{equation}
\psi_n(\mbf{r}) 
=e^{i\mbf{k}\cdot\mbf{r}}\delta_{n,0} -\frac{m}{4\pi}\frac{e^{ik_n r}}{r} \sum\limits_{s=-\infty}^\infty \langle k_n\mbf{\hat{r}}|V_{n-s}|\psi_s\rangle,
\label{eq:asymptoticWF}
\end{equation}
where $\mbf{r}$ is the separation vector between the scatterers. From the asymptotic form of the wavefunction, we see that the scattering amplitude for a transition from the initial state to the $n^{\mathrm{th}}$ Floquet mode is
\begin{equation}
f_n(k_n\mbf{\hat{r}},\mbf{k})=-\frac{m}{4\pi}\sum\limits_{s=-\infty}^\infty \langle k_n\mbf{\hat{r}}|V_{n-s}|\psi_s\rangle.
\label{eq:deff}
\end{equation}
By premultiplying Eq.~\eqref{eq:LSEPsi} with $-(m/4\pi)\langle \mbf{k}_n|V_{i-n}$ and summing over $n$ we find (after relabeling indices) an integral equation for the scattering amplitudes:
\begin{align}
f_n(\mbf{p}_n,\mbf{k})
&= -\frac{m}{4\pi}\langle \mbf{p}_n|V_n|\mbf{k}\rangle +\sum\limits_{s=-\infty}^\infty \int\dvec{q}\nonumber \\
&\times
\frac{\langle \mbf{p}_n|V_{n-s}|\mbf{q}\rangle}{k^2/m-q^2/m+s\omega + i\epsilon} f_s(\mbf{q},\mbf{k}).
\label{eq:intEqnf}
\end{align}
This system can be solved for $f_n(\mbf{p}_n,\mbf{k})$ by first solving the similar integral equation for $f_n(\mbf{p},\mbf{k})$ and then inserting the result into the right side of Eq.~\eqref{eq:intEqnf}. The physical amplitudes in Eq.~\eqref{eq:deff} are then obtained by setting $\mbf{p}_n=k_n\hat{\mbf{r}}$ where $\mbf{k}$ is the relative momentum of the incoming particles.

{\it Zero-range solution.---} We now consider s-wave scattering from a zero-range, time-periodic potential where the resonance molecule is a shallow dimer with binding energy $E_D=1/(m\abar^2)$ where $\abar= a(\Bbar)$. As discussed in \cite{Langmack:prl:2015}, the effects of an oscillating magnetic field under these conditions can be expressed in terms of an effective field theory containing a contact interaction with time-dependent scattering length $a(t)\equiv a(B(t))$ from Eqs.~\eqref{eq:feshbach} and \eqref{eq:BofT}. Using this effective field theory, the matrix elements $\langle \mbf{p}|V(t)|\mbf{k}\rangle$ evaluate to $g(t)/m$ where $g(t)=4\pi/(1/a(t)-2\Lambda/\pi)$ and $\Lambda$ is a cutoff in the magnitude of the momenta of virtual particles. $g(t)$ can be expanded in Fourier modes: $g(t)=\sum_{n} g_n \exp(-in\omega t)$. From this we conclude that $\langle \mbf{p}|V_n|\mbf{k}\rangle=g_n/m$. It should be noted that the couplings $g_n$ have no dependence on the momenta $\mbf{k}$ and $\mbf{p}$. This implies that the scattering amplitudes $f_n(\mbf{p},\mbf{k})$ only depend on the magnitude of $\mbf{k}$, allowing us to relabel the amplitudes as $f_n(k)$. Eq.~\eqref{eq:intEqnf} then simplifies to
\begin{align}
\sum\limits_{s=-\infty}^\infty\left[\delta_{n,s}+\frac{g_{n-s}}{4\pi} \left(ik_s+\frac{2\Lambda}{\pi}\right)\right] f_s(k)
&= -\frac{g_n}{4\pi}.
\label{eq:intEqnfZeroRange}
\end{align}
We can reproduce the time-independent case by setting $\Btil=0$. This corresponds to replacing $g(t)$ with $\bar g=4\pi/(1/\abar-2\Lambda/\pi)$ which implies that $g_0=\bar g$ and $g_n=0$ for $n\neq 0$. Inserting these results into Eq.~\eqref{eq:intEqnfZeroRange}, we find $f_0(k)=1/(-1/\abar-ik)$ and $f_n(k)=0$ for $n\neq 0$ as we would expect for zero-range scattering from a time-independent potential with scattering length $\abar$.

To treat the case where $1/a(t)$ has small deviations from $1/\abar$, we expand $g(t)$ in powers of $1/a(t)-1/\abar$. Following Ref.~\cite{Mohapatra:pra:2015}, we drop terms of order $(1/a(t)-1/\abar)^2$ and higher because these terms are suppressed by higher powers of $1/\Lambda$. We then expand $1/a(t)-1/\abar$ in powers of the dimensionless magnetic field variable $\btil = [a'(\Bbar)/\abar]\Btil$ which for $\Bbar$ near $B_0$ approaches $-\Btil/(\Bbar-B_0)$. Keeping terms up to order $\btil^2$ we find
\begin{equation}
g(t)=\bar g + \frac{\bar g^2}{4\pi\abar}\left(\btil\cos(\omega t) +\frac{\Bbar-B_0}{\Delta}\btil^2\cos^2(\omega t)\right).
\label{eq:gexp}
\end{equation}
Since we consider the case where the resonance molecule is a shallow dimer controlled  via a MFR, we treat $(\Bbar-B_0)/\Delta$ as a small parameter. We therefore omit the second term in Eq.~\eqref{eq:gexp}. Under these conditions, the Fourier components of $g(t)$ are 
\begin{align}
g_0 =\bar g,~~~
g_{\pm 1} = \bar g^2/(8\pi\bar a)\,\btil, 
\label{eq:gns}
\end{align}
with all other components equal to zero. The resulting solutions to Eq.~\eqref{eq:intEqnfZeroRange} include the effects of the oscillating potential up to second order in $\btil$ and zeroth order in $(\Bbar-B_0)/\Delta$. Since the coupling between Floquet channels is suppressed by powers of $\btil$, the sum in Eq.~\eqref{eq:intEqnfZeroRange} can be truncated to a finite range of Floquet modes. On the other hand, for large $\btil$ or very small $\omega$ all Floquet modes contribute, and Eqs.~\eqref{eq:intEqnfZeroRange} cannot be solved by truncating the sum. This signals the breakdown of our assumption that we can describe the scattering process in terms an incoming momentum eigenstate coupled to a discrete spectrum of outgoing momentum eigenstates.

\begin{figure} [t]
\centering
\includegraphics[width=\columnwidth]{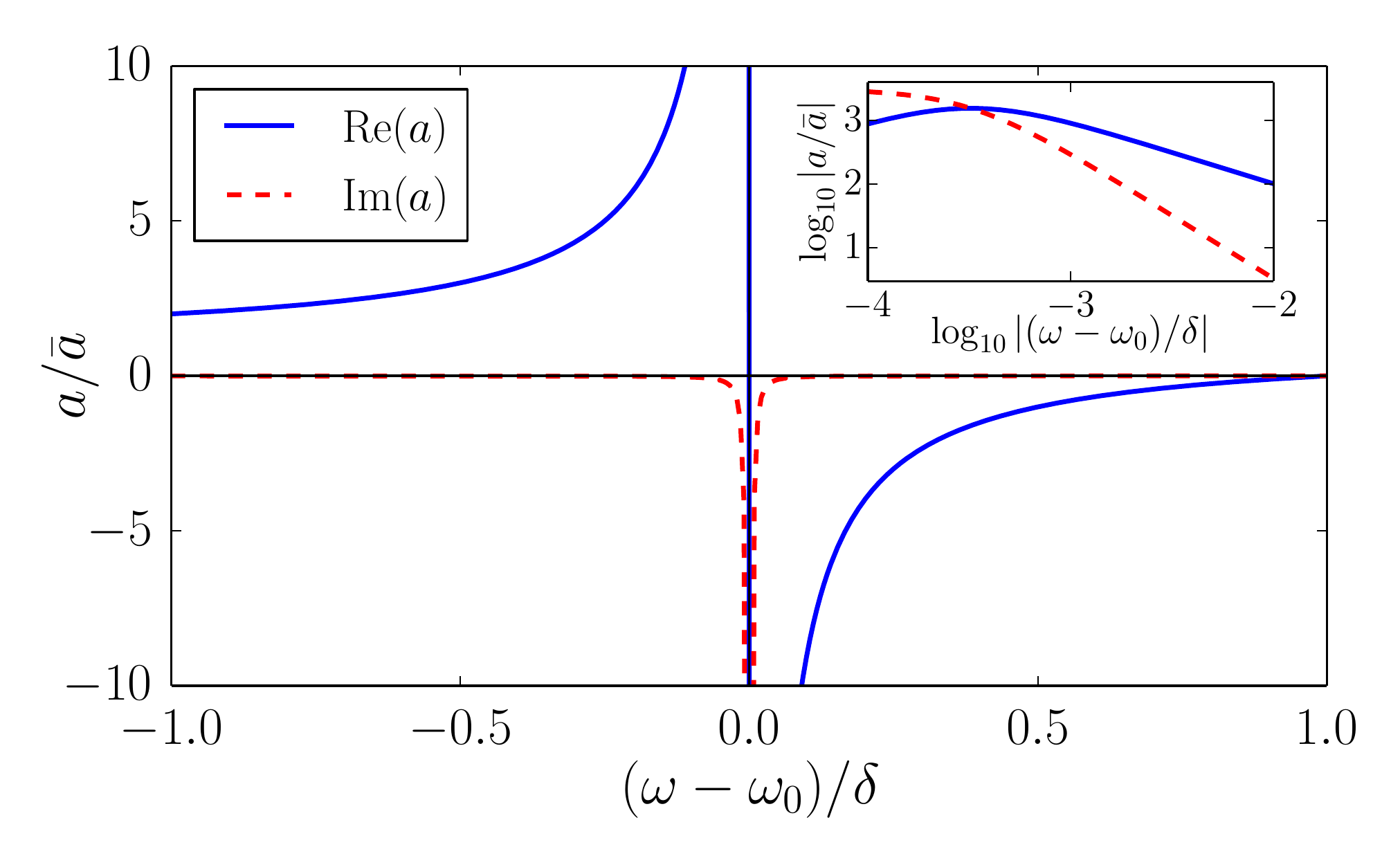}
\caption{The real and imaginary parts of $a$ as functions of $\omega$ for a MMFR. Here we have chosen $\btil=0.05$. The inset shows the absolute values of the same quantities on a logarithmic scale.}
\label{fig:scatLengthZoomed}
\end{figure}

{\it Frequency dependent scattering length.---}
We now extract the function $a(\omega)$ in Eq.~\eqref{eq:resonance_eqn} by noting that $a(\omega)=-\lim_{k\rightarrow 0}f_0(k)$, where $f_0(k)$ is determined by Eq.~\eqref{eq:intEqnfZeroRange} with the couplings in Eq.~\eqref{eq:gns}. For $\btil^2\ll 1$ it is sufficient to truncate the sum in Eq.~\eqref{eq:intEqnfZeroRange} to the five Floquet modes from $n=-2$ to $2$. After truncation, the system of equations can be solved analytically. Taking the limits $\Lambda\rightarrow \infty$ and $k\rightarrow 0$, we find for $a(\omega)$ a complicated analytic expression that depends upon $\abar$, $E_D$, $\btil^2$, and $\omega$. For $\omega$ near $E_D$ and at order $\btil^2$, this analytic expression is equivalent to Eq.~\eqref{eq:resonance_eqn} with the resonance parameters
\begin{align}\label{eq:scaling_behavior}
\delta/\omb &= (1/2)\,\btil^2, \nonumber \\
\Delta\om_0/\omb &= (\sqrt{2}/2)\,\btil^2, \nonumber \\
\gamma\abar &= (1/8)\,\btil^2.
\end{align}
The quadratic scaling of $\delta$ and $\Delta\om_0$ with the amplitude of the oscillating field agree with those in OFR \cite{Julienne_1999} and rf/mwFR \cite{Schmiedmayer_rfFR_2010,Dalibard_mwFR_2010}. The scaling law for $\gamma$ has not been reported in previous work on OFR or rf/mwFR. Figure \ref{fig:scatLengthZoomed} plots the real and imaginary parts of the scattering length as functions of the frequency, demonstrating the familiar Feshbach resonance shape. Molecule formation limits the maximum value of $\mathrm{Re}(a(\omega))$. The inset of Fig.~\ref{fig:scatLengthZoomed} emphasizes this point by plotting the real and imaginary parts of $a(\omega)$ on a logarithmic scale. 

\begin{figure} [t]
\centering
\includegraphics[width=\columnwidth]{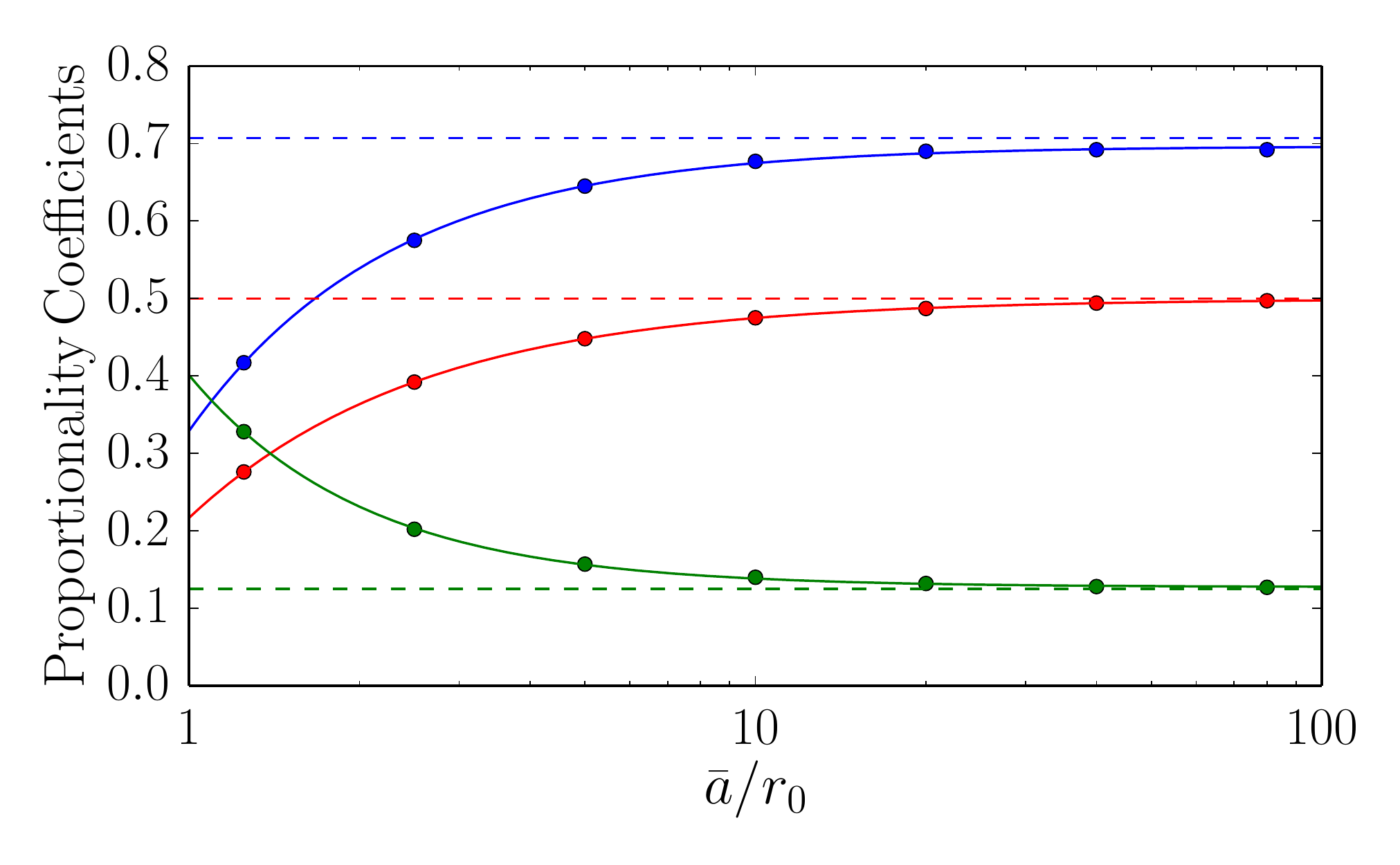}
\vspace{-20px}
\caption{Numerical results from the square-well model for the coefficients of $\btil^2$ for the dimensionless resonance parameters $\Delta\om_0/\omb$, $\delta/\omb$, and $\gamma\abar$ (top to bottom, solid curves) as functions of $\abar$ divided by the range of the potential $r_0$ along with power-law fits to guide the eye. In the large $\abar/r_0$ limit, the coefficients approach the universal numbers given in Eqs.~\eqref{eq:scaling_behavior} (dashed horizontal lines).} 
\label{fig:universallimit}
\end{figure}

We check these results by explicitly calculating the scattering wave function for a square-well potential with oscillating depth following the methods outlined in Ref.~\cite{Reichl:prb:1999}. The parameters of the potential are tuned to reproduce the binding energy of the shallow dimer and the shift of the shallow dimer due to a small shift in the magnetic field. The function $a(\omega)$ for this model is well-parametrized by Eq.~\eqref{eq:resonance_eqn}. In the zero-range limit, the resonance parameters are $\delta/\omb = 0.50\,\btil^2$, $\Delta\om_0/\omb = 0.69\,\btil^2$, and $\gamma\abar = 0.13\,\btil^2$. These results agree quantitatively with those in Eq.~\eqref{eq:scaling_behavior}. Figure~\ref{fig:universallimit} shows the convergence of the resonance parameters in the square-well model to the universal, zero-range results.

{\it Discussion.---} The MMFR mechanism can be realized in experiment by producing a sample of atoms with one hyperfine scattering channel, selecting a resonance molecule either in the scattering channel or else in another hyperfine channel, and then oscillating the magnetic field with a frequency near the transition frequency $\omb$ of the molecule and with an adjustable amplitude $\Btil$. Near the resonance, $a(\omega)$ can be parametrized by Eq.~\eqref{eq:resonance_eqn}. The resonance parameters $\delta$, $\Delta\om_0$, and $\gamma$ scale as $\Btil^2$ with coefficients that could be either calculated by solving the time-dependent Schr{\"o}dinger equation with accurate microscopic potentials or else measured experimentally. If the resonance molecule is a shallow dimer in the scattering channel, the dimensionless resonance parameters have the universal values in Eq.~\eqref{eq:scaling_behavior}. 

We illustrate the universal results in Eq.~\eqref{eq:scaling_behavior} with $^7$Li atoms near the broad MFR at $B_0=737.69(12)$~G. We consider the conditions of the wiggle spectroscopy experiment in Ref.~\cite{Hulet1302}: $\Bbar=734.5$~G and $\Btil=0.57$~G, which implies $\btil=-0.18$. Inserting the MFR parameters reported in Ref.~\cite{Hulet1302} into Eq.~\eqref{eq:feshbach} gives $\abar = 1100~a_0$ which implies $\omb=450$~kHz. The universal results for the MMFR parameters are $\delta=7.5~$kHz, $\Delta\om_0=10~$kHz, and $\gamma=1/(2.6\times 10^5 a_0)$. To utilize this MMFR, the frequency resolution must be significantly finer than $\delta$. The lifetime of the resonance molecule in the presence of the oscillating field is $1/(2\gamma\abar\delta)=16$~ms. The maximum value of $|\re{a}|$, which occurs at the detunings $|\om-\om_0|=\gamma\abar\delta=32~$Hz, is $1/2\gamma=1.3\times 10^5 a_0$. This is more than a factor of $100$ enhancement over $\abar$. These extrema occur at the frequencies where $|\re{a}|=|\im{a}|$ (see inset of Fig.~\ref{fig:scatLengthZoomed}). Larger values of the detuning are required in order to have $|\re{a}| \gg |\im{a}|$. For example, when $\om-\om_0=-0.5\,\delta$, $\re{a}/\abar=3.0$ and $\im{a}/\abar=-0.039$. Since $|\im{a}/\re{a}|\approx\gamma\abar\delta/|\om-\om_0|$ is proportional to $\btil^4$, $|\im{a}|$ can be made much smaller than $|\re{a}|$ by decreasing $\btil$. This necessitates finer frequency resolution, since $\delta$ also scales as $\btil^2$. Since $\im{a}$ diminishes more rapidly with $\tilde b$ than $\delta$, MMFR can significantly enhance $\re{a}$ without introducing catastrophic losses from $\im{a}$.  

{\it Summary.---} We have examined a new mechanism for resonantly enhancing the s-wave scattering length $a$ by tuning the frequency of an oscillating magnetic field parallel to the spin-quantization axis to near the binding frequency of a molecule. Along with enhancing the real part of $a$, the oscillating field also generates an imaginary part associated with molecule formation. The real and imaginary parts of $a$ can be controlled by the frequency and amplitude of the oscillating magnetic field. In the case that the resonance molecule is a shallow bound state in the scattering channel, the zero-range dimensionless resonance parameters are universal functions of $\btil$. Using these universal results, we demonstrated that MMFR can significantly enhance the scattering length without introducing catastrophic losses.

\begin{acknowledgments}
This research was supported by the National Science Foundation under grant PHY-1310862. Many thanks are due to Dr.~Cheng Chin whose insightful question sparked this research topic. I thank Dr.~Artem Volosniev and Logan Clark for helpful discussions. I also thank Dr.~Eric Braaten for invaluable feedback during the writing of this letter.
\end{acknowledgments}

\end{document}